\documentclass[lettersize,journal]{IEEEtran}

\usepackage{amsmath,amsfonts}
\usepackage{stmaryrd}
\usepackage{algorithm}
\usepackage{algpseudocode}
\usepackage{algorithmicx}
\usepackage{array}
\usepackage{textcomp}
\usepackage{stfloats}
\usepackage{url}
\usepackage{verbatim}
\usepackage{graphicx}
\usepackage{xspace}
\usepackage{bbold}
\usepackage{multirow}
\usepackage[table,svgnames]{xcolor} 
\usepackage{braket}

\usepackage{amsmath}
\usepackage{amssymb}

\usepackage{tikz}
\usetikzlibrary {arrows.meta}
\usetikzlibrary {shapes.geometric}
\usetikzlibrary {positioning}

\usepackage{booktabs}
\usepackage{marginnote}

\usepackage[most]{tcolorbox}

\usepackage{cite}

\DeclareMathOperator*{\argmax}{arg\,max}

\algnewcommand{\LineComment}[1]{\State \(\triangleright\) #1}

\newtheorem{example}{Example}
\newtheorem{remark}{Remark}

\tikzset{
    simple_node/.style={
        draw=black,
        align=left,
        minimum height=2.5em,
        rounded corners=2mm,
        align=center
    },
    classifier_node/.style={
        trapezium,
        trapezium angle=70,
        text width=50.0,
        draw=black,
        align=center,
        minimum height=2.5em,
        align=center
    },
    simple_line/.style={
        draw=black,
        -{Stealth[length=2.5mm]}
    }
}
\definecolor{color_1}{RGB}{36,160,216}
\definecolor{color_2}{RGB}{216,84,151}
\definecolor{color_3}{RGB}{176,201,43}
\definecolor{color_4}{RGB}{101,101,108}

\begin{document}

\title{Physics-Informed Detection of Friction Anomalies\\in Satellite Reaction Wheels}

\author{Alejandro Penacho Riveiros, Nicola Bastianello, \emph{Member}, \emph{IEEE}, Karl H. Johansson, \emph{Fellow}, \emph{IEEE}, and Matthieu Barreau
\thanks{Paper submitted on January 26, 2026. This work was supported by the Ultimate project of the European Union’s Horizon Research and Innovation Actions program under grant agreement No. 101070162, and by the Wallenberg AI, Autonomous Systems and Software Program (WASP) funded by the Knut and Alice Wallenberg Foundation. Computation resources were provided by the National Academic Infrastructure for Supercomputing in Sweden (NAISS), partially funded by the Swedish Research Council through grant agreement no. 2022-06725.}%
\thanks{A. Penacho Riveiros, N. Bastianello, K. H. Johansson, and M. Barreau are with the Department of Decision and Control Systems, Digital Futures, KTH Royal Institute of Technology, Stockholm, 100 44 Sweden {\tt\small\{alejpr, nicolba, kallej, barreau\}@kth.se}}
}


\markboth{Journal of \LaTeX\ Class Files,~Vol.~14, No.~8, August~2021}%
{Shell \MakeLowercase{\textit{et al.}}: A Sample Article Using IEEEtran.cls for IEEE Journals}


\maketitle

\begin{abstract}
As the number of satellites in orbit has increased exponentially in recent years, ensuring their correct functionality has started to require automated methods to decrease human workload. In this paper, we present an algorithm that analyzes the on-board data related to friction from the Reaction Wheel Assemblies (RWAs) of a satellite and determines their operating status, distinguishing between nominal status and several possible anomalies that require preventive measures to be taken.
The algorithm is based on a hybrid system model and combines changepoint detection, dynamic programming, and maximum likelihood estimation. A classifier then uses the extracted information to determine the status of the RWA. The classifier is trained on a labelled dataset produced by a high-fidelity simulator, comprised for the most part of nominal data.
The algorithm combines model-based and data-based approaches to obtain satisfactory results with an accuracy around $\mathbf{95\boldsymbol{\%}}$, a false positive rate below $\mathbf{3\boldsymbol{\%}}$, and an execution time in the order of a tenth of a second, making it suitable to run on-board. Moreover, the algorithm allows for easy tuning and explainability of the results obtained.
\end{abstract}


\section{Introduction}

\subsection{Motivation}
As satellites orbiting Earth have become more frequent during the last decades \cite{mcdowell_general_2023}, the need to maintain their correct functioning has become a major concern of space agencies. Often, satellite repairs are unfeasible, so failures usually lead to the termination of the mission. Even when repair is possible, it tends to be very expensive, as was the case of the installation of a new camera for the Hubble telescope lens, which required an entire mission \cite{fricke_sts-61_1994}. This makes the capability of detecting impending new failures a fundamental component of any satellite system.

Among the components that can fail in a satellite, the Reaction Wheel Assemblies (RWAs) are of particular interest.
RWAs are fundamental components of the Attitude and Orbit Control System (AOCS). Their purpose is to exert a torque on the body of the satellite to control its attitude. Rather than using action--reaction principles as a Reaction Control System (RCS) does, an RWA uses angular momentum conservation to achieve greater accuracy with no propellant consumption, which is highly advantageous for star tracking and long-duration missions.

\begin{figure}
    \centering
    \includegraphics[width=\linewidth]{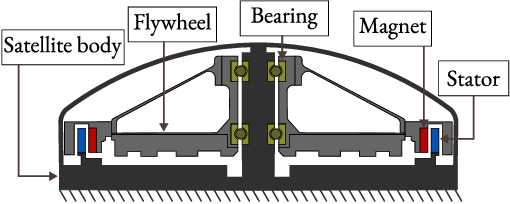}
    \caption{Cross-section of an RWA, adapted from \cite{pantaleoni_curing_2014}.}
    \label{fig:rwa-diagram}
\end{figure}

The operating principle of RWAs is fairly simple. Inside the assembly, a flywheel is kept rotating permanently, see Figure~\ref{fig:rwa-diagram}. When there is a need to rotate the satellite in one direction, the flywheel is accelerated in the opposite direction using an electric motor. Due to the conservation of angular momentum, this produces the desired torque in the main body of the satellite \cite{narkiewicz_generic_2020}.
To control the three axes of rotation of the satellite, at least three RWAs need to be mounted. However, to achieve redundancy and flexibility in the speed of rotation of the flywheels, it is common to install four RWAs\cite{brown_-flight_2012}.

As an RWA is a mechanical component in almost permanent motion, it is more prone to failure than other satellite parts. Mechanical degradation, vibrations, and wear occur, compromising its integrity \cite{kirsch_cage_2012}. Particularly, the bearings, where contact between the rotating disc and the non-moving satellite body takes place, are home to several complex phenomena involving elastohydrodynamics (interaction between a moving fluid, such as lubricant, and a deformable solid material). These phenomena cause sudden changes in friction at the bearings that make the wheel modeling particularly challenging.
A failure of an RWA might imply losing satellite functionality and can seriously compromise the mission. 
This was the case in space missions Kepler \cite{larson_kepler_2014}, Dawn \cite{rayman_dawns_2014-1}, and SDO \cite{ekinci_solar_2014}, in which the capabilities of the satellites were compromised after RWA failures \cite{sahnow_operations_2006, ekinci_solar_2014}.

Although there have been attempts at explaining the causes of RWA failures \cite{bialke_newly_2017}, the complexity of the bearing inner dynamics makes it hard to determine the precise condition of the component. Additionally, limited number of RWA sensors implies that a complete monitoring of the situation is impossible.
Even though the nature of these failures remains unknown, recent studies have shown that they are usually preceded by changes in the behavior of the friction at the bearings, which can be observed using on-board sensors \cite{kampmeier_reaction_2018}. Such anomalies are usually caused by RWA changes, such as alterations of the lubricant or an increase in the contact surface in the bearings.
It is of interest to detect these changes as they can be used to anticipate possible future failures of the wheel and take measures to improve the health of the RWA \cite{pantaleoni_curing_2014}. However, not all changes imply a future failure, so it is necessary to distinguish between different types of anomalous behavior to provide the appropriate warnings. Therefore, it is not enough to perform anomaly detection, but categorization should also be considered.

\subsection{Contribution}
In this paper, we develop a method that analyzes the on-board data collected from an RWA during a window of time of a few hours to determine whether anomalies are present and, if positive, which ones. The algorithm can be implemented onboard the satellite thanks to its low computation cost. We make the following specific contributions:
\begin{itemize}
    \item We derive a novel friction model that incorporates sudden friction variations due to quick changes in the internal geometry of the system. The model represents their statistical properties, which can be obtained experimentally.
    \item An end-to-end friction detection algorithm is derived that takes a window of a few hours of measurements by on-board sensors to determine whether the observed friction characteristics of the RWA belong to nominal or anomalous conditions. The method combines physical model knowledge with data to provide an accurate diagnosis procedure.
    \item A modification in the changepoint classification algorithm that accounts for the possibility of false negatives, making the algorithm robust to false positives in the detection of changepoints.
    \item The algorithm and each of its components are experimentally evaluated using high-fidelity simulation data from Thales Alenia Space (TAS) \footnote{https://www.thalesaleniaspace.com}, with no prior knowledge of the simulation model.
\end{itemize}

\subsection{Related work}
Anomaly and fault detection in satellites has received considerable interest in recent years, primarily due to their proliferation and the need for constant monitoring. Research done in this direction has attempted to produce algorithms capable of analyzing all the information coming from the satellite to determine a wide variety of possibly faulty conditions.
The high availability of data and the lack of models for the complete satellite has led to almost exclusively data-driven approaches, using tools such as  Support Vector Machines (SVM) \cite{fuertes_improving_2016}, Relevance Vector Machines (RVM) \cite{fujimaki_anomaly_2005}, Principal Component Analysis (PCA) \cite{gao_fault_2012}, or LSTM networks \cite{ibrahim_machine_2019}.
These ideas have led to algorithms used in actual space missions, like Orca \cite{bay_mining_2003} in the International Space Station and the IMS tool \cite{iverson_inductive_2004} in the Space shuttle.

Satellite-specific methods are very versatile as they can catch a wide range of possible malfunctions in the vehicle. However, they consider the vehicle as a whole, ignoring the intricacies of each component. This makes them less sensitive to subtle changes in the behavior of any of these individual components.


It is then reasonable to find detection methods specific to the component of interest.
Focusing on RWA, the simplicity and generally well-understood dynamics of these devices have allowed the development of more specialized techniques capable of leveraging expert knowledge. This has led to extremely detailed models to predict vibrations in the RWA \cite{masterson_development_2002, longato_microvibration_2023}. Although accurate, such models are unsuitable for anomaly detection during space missions, as they usually involve signals not available outside ground tests.

The problem of fault and anomaly detection for RWAs has been analyzed mainly through machine learning \cite{bellali_parameter_2012, abdel_aziz_efficient_2024}. Although these methods have been effective in their purpose, they consider the behavior of the electric motor rather than focusing on the friction torque, so they are not so precise in determining abnormal behavior of the latter. This was done only recently using unsupervised machine learning \cite{naik_using_2020}, but considering only normal and abnormal behavior, that is, without isolation of the anomaly. Although research have been done considering different possible anomalies, their are limited to increases in either dry or viscous friction \cite{riveiros_real-time_2024}.


Closely related to RWAs, rotary machinery analysis is a richer field of literature. Although purely model-based approaches have been proposed in the past \cite{loparo_fault_2000}, they are usually focused on alterations in the vibration patterns of the machine. Most of the research is based on data-driven techniques, capable of assessing a wider variety of faulty conditions usually related to problems in the bearings \cite{hamadache_comprehensive_2019}.

Most of the rotary machinery analysis techniques follow a common approach: first, some signals are measured from the machine (usually displacement or acceleration), then features are extracted (statistical ones like mean and variance, or frequency-related like Power Spectral Density), and finally, some machine learning methods are used to determine the status of the machine. Such methods include nearest neighbor \cite{mechefske_fault_1992}, neural networks \cite{li_detection_1998}, SVM \cite{sugumaran_effect_2011}, linear regressions \cite{rocchi_fault_2014}, or Broad Learning Systems (BLS) \cite{liu_fault_2023}, among others.

These methods have achieved important improvements in recent years by incorporating knowledge of a model of the machine, leading what is known as hybrid methods \cite{leturiondo_validation_2017}. A popular technique is to fuse these two sources of knowledge through Physics Informed Machine Learning (PIML), which has experienced a rapid development in the last years \cite{shen_machine_2023}.

Although evident progress has been achieved as new features and classification methods have been developed, their focus on the vibrational behavior of the machine makes them unsuitable for satellites, where this behavior is not of much interest. Regarding friction-related research, most of the interest is on the low-speed behavior, where complex phenomena like the Stribeck curve or stick-slip occur \cite{popovic_experimental_2003, geissmann_switching_2024}.
Research on switching behavior in friction has also received some attention, but restricted to deterministic systems\cite{marino_switching_2023}.

In conclusion, there has not been any work in the study of friction systems (particularly RWAs) with random jumping friction behavior, which poses unique challenges for detecting anomalies. This article aims to fill the existing gap in the literature by providing a hybrid model-data-based anomaly detection and isolation algorithm.

\subsection{Outline}
The paper is structured as follows: Section II states the problem. In Section III, we introduce the RWA model used for the algorithm design. Section IV develops the algorithm. In Section V, we comment on the tuning of the algorithm's parameters. Section VI describes the experimental evaluation and illustrates the performance of the algorithm.. Finally, we provide conclusions and possible future work in Section VII.
\label{sec:introduction}

\section{Problem Formulation}
We study the problem of determining if the friction torque of an RWA has an anomalous behavior based on a sequence of measurements taken by its sensor system at fixed time intervals.
Each measurement at time $t_k \in \mathbb{R}$, where $k \in \mathbb{N}$ is the index of the measurement, consists of the spin rate $\omega_k$ of the RWA, the voltage $V_k$, and the current $I_k$ in the motor that accelerates (or deccelerates) the RWA. Using conservation of angular momentum, energy equations at the motor, and filtering techniques, we obtain an unbiased estimation of the friction torque $f_k$ at the bearings of the RWA using the estimator
\begin{align}
    \hat{f}_k &= J \frac{\omega_{k+1} - \omega_{k-1}}{t_{k+1}-t_{k-1}} - K_T I_k,
    \label{eq:all_signals}%
\end{align}
which follows from
\begin{align}
    J \dot{\omega} &= T_\text{motor} + f = K_T I + f
\end{align}
where $T_\text{motor}$ is the torque produced by the motor, $J$ is the wheel inertia, and $K_T$ is the motor torque constant. The estimated $\hat{f}_k$ can be filtered to remove the effect of noise. In this paper, we do not consider the details of this filtering process, but rather assume that $\hat{f}_k = f_k + v_k$, where $v_k$ is random noise with zero mean, as the filter is expected to be unbiased.

Since we are interested only in the friction characteristics of the RWA, the input to our algorithm is a sequence of spin rate and estimated friction torque pairs:
\begin{align}
    X = \{(\omega_k, \hat{f}_k)\}_{k=1}^N.
\end{align}
This sequence is normalized using average values of historical data, so the variables take values of order one.
All indices used in the paper are given in Table~\ref{tab:indexes}. Subscripts and superscripts that are part of the name of the variable are written upright (like $N_\mathrm{cp}$) and those that are indices are written in italics (like $\omega_k$).

\begin{table}[H]
    \centering
    \begin{tabular}{c p{35mm} c}
        \toprule
        \textbf{Index} & \textbf{Meaning} & \textbf{Range}
        \\
        \midrule
        $\ell$ & Datapoint in dataset & [1, $N_\mathrm{D}$]
        \\
        $k$ & Measurement in datapoint & $[1, N_\mathrm{m}]$
        \\
        $i$ & Changepoint & $[1, N_\mathrm{cp}]$
        \\
        $s$ & Switching system & $[1, N_\mathrm{s}]$
        \\
        \bottomrule
        \addlinespace
    \end{tabular}
    \caption{Indexes and their range}
    \label{tab:indexes}
\end{table}

Consider we have $N_D$ sequences of measurements $X_\ell$, indexed by $\ell = [1, N_D]$. Each of these sequences is labeled with an RWA status $\theta_\ell$. We can collect these sequences and their corresponding labels into a dataset $D = \{(X_\ell, \theta_\ell)\}_{\ell=1}^{N_\mathrm{D}}$. This dataset can be obtained from ground tests, in-orbit measurements or, as in our cause, high-fidelity simulators.

The status of a RWA is related to its friction properties. In the dataset $D$, each sequence $X_\ell$ is labeled by the status $\theta_\ell$ of the RWA from which that sequence of measurements was taken. An RWA can be affected by $N_\mathrm{an}$ different anomalies, with two rules: anomalies are either active or inactive, and any number of anomalies can be active in an RWA simulteneously. This is modeled by defining the set of possible status as $\Theta = \{0, 1\}^{N_\mathrm{an}}$, such that each element of $\theta$ indicates whether the corresponding anomaly is inactive ($0$) or active ($1$). The objective of the algorithm is, then, to determine for a new sequence $X$ what is the status of the RWA that generated it, that is, what anomalies affect it.

\begin{figure}
    \centering
    \includegraphics[width=1.0\linewidth]{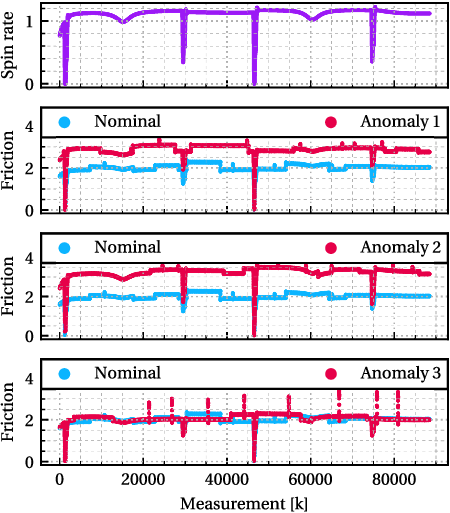}
    \caption{Examples of runs labeled with different anomalies. For each case, the anomalous run is shown in red while a nominal run is plotted in blue to help compare them.}
    \label{fig:example_ds}
\end{figure}

We evaluate our approach on data generated by a high-fidelity simulator developed by TAS, similar to the one used in \cite{flammini_meteosat_2024}. To illustrate how this dataset looks like depending on the anomaly, we show three examples in Figure~\ref{fig:example_ds}.
Some early observations indicate that, even in the nominal case, there are sudden variations of the friction independent of the spin rate (for example, the one close to measurement $4000$ in the nominal case). Sudden changes in the bearings cause such variations due to, for example, oil accumulations in small pockets. Although their presence does not necessarily imply RWA anomalies, they must be considered, since excessive friction changes due to these variations are usually concerning. A model that captures the behavior of an RWA must, therefore, include these sudden friction changes.

Regarding differences between anomalies, Anomaly~1 produces a constant increase of friction torque at all spin rates, while Anomaly~2 increases friction torque proportionally with the spin rate. Additional anomalies relate to friction differences produced in the sudden events described in the previous paragraph.

\textbf{Problem Statement:} Develop an algorithm that can detect the status $\theta$ of an RWA based on a sequence of measurements $X$ of spin rate and estimated friction torque, and train it based on historical labeled data given as a dataset $D$.



\label{sec:formulation}

\section{RWA Modeling}\label{sec:model}
To solve the problem formulated in the previous section, we propose a hybrid approach combining data-based and model-based detection methods. In this section, we derive the model used for the model-based part, developed from a combination of dataset observation, knowledge of the RWA mechanism, and literature on the topic.

\subsection{Friction Model}

We model the RWA as a friction system:
\begin{equation}
    f_k = f_k^\mathrm{d} \text{sign}(\omega_k) + f^\mathrm{v} \omega_k + v_k \label{eq:model_friction}
\end{equation}
where \eqref{eq:model_friction} corresponds to a classical dry and viscous friction model with additive Gaussian noise $v$ \cite{armstrong-helouvry_control_1991}. The dry and viscous friction coefficients are denoted $f_k^\mathrm{d}$ and $f^\mathrm{v}$, respectively, where the dry coefficient can change with time.

Each component of \eqref{eq:model_friction} models a different friction factor.
Starting from the end, the Gaussian noise $v$ has three sources: randomness in the motion of the bearing and the contact inside it, sensor noise, and inaccuracies in the friction torque estimation. 
Among all possible noise models available, the Gaussian one was chosen for its theoretical properties and its popularity in the literature \cite{gustafsson_slip-based_1997, carrara_estimating_2013, hacker_reaction_nodate}, as well as for its good fit with the observed data.

The viscous friction, on the other hand, is produced by the lubricant inside the bearing \cite{lee_-flight_2015}. This parameter is known to be affected by the temperature and the pressure of the lubricant, as well as its age \cite{seeton_viscositytemperature_2006}. However, these variations take place over longer time intervals than we analyze; so we can assume this coefficient to be constant for our algorithm. Note that, although constant, this coefficient can still be affected by an anomaly that increases its value permanently.

Finally, the dry friction models the contact between solid surfaces, and is given by
\begin{equation}
    f_k^\mathrm{d} = f^{\bar{\mathrm{d}}} + \sum_{s=1}^{N_\textrm{s}} f_k^s. \label{eq:model_dry-friction}
\end{equation}
In the RWA bearings, this contact occurs between the the inner and outer races of the bearing and the balls, and between the balls themselves \cite{longato_microvibration_2023}. The contact surface at every instant is not necessarily constant, which can lead to changes in the dry friction coefficient. 
To represent these changes, we define in \eqref{eq:model_dry-friction} the dry friction coefficient as the sum of several terms, the first of which is the constant baseline dry friction $f^{\bar{d}}$ that remains constant at all times:
The remaining components $f_k^s$ capture the effect of variations in the contact surface of the bearings. The evolution of each of these components is described by what we refer to as a Friction Switching System (FSS). The motivation behind using several FSS in the model of a single RWA is to capture independent phenomena with different properties. Such phenomena can range from concentrations of pockets of oil (well reported in the literature \cite{lee_-flight_2015} to displacements of the different components of the bearing (observed in the past but not well characterized so far \cite{kirsch_cage_2012}), which may have very different time scales. By modeling them with different FSS, we can also distinguish anomalies affecting one or another phenomenon, enhancing the accuracy of the categorization of anomalies. An example of using several FSS in a single RWA model is given in Example~\ref{ex:hmm}.


\subsection{FSS} \label{sec:fss}
We define an FSS as a Hidden semi-Markov Model (HSMM) \cite{yu_hidden_2010} with state $q \in \left[ 1, q_\text{max} \right] \subset \mathbb{N}^{\geq 0}$, the latter being the set of natural numbers including $0$.
An HSMM is an extension of the standard HMM, in which a duration $\tau \in \mathbb{N}^{\geq 0}$ is associated to the state of the system. The duration indicates the number of steps the HSMM stays in that state before the next transition. Every time there is a state transition, a new duration $\tau'$ is drawn from a random distribution.

To formally define the HSMM we need a transition function and an observation function. The transition functions specifies how likely the system is to transition to state $q$ for a duration $\tau'$ after having spent $\tau$ steps in state $q$. It is defined by the conditional probability function $P(q', \tau' | q, \tau)$.

For the observation function, we propose a simpler alternative to the standard HSMM formulation:
when a transition takes place at step $k_\mathrm{tr}$, a friction value $f_{k_\mathrm{tr}}$ is drawn from a probability function given by $P^\mathrm{f}(f_{k_\mathrm{tr}}|q)$. Between transitions, the observed friction value produced by the FSS is constant, and equal to the friction $f$ generated in the last transition, that is, $f_{k \in [k_\mathrm{tr}, k_\mathrm{tr} + \tau)} = f_{k_\mathrm{tr}}$.

The FSS emulates unobservable phenomena inside the bearing of the RWA, such as the distribution of lubricant, the relative position of the inner and outer races, or any other aspect that may affect the dry friction. The state variable $q$ is the current configuration of the system and $f$ is the generated friction. The configuration represents, for instance, various stable positions of the bearings, or whether there is oil in a certain gap in the bearings or not.

Since we only use the FSS for modeling these friction phenomena, we impose some restrictions on their behavior based on observations from the TAS dataset and previous experience with friction in RWAs \cite{lee_-flight_2015, pantaleoni_curing_2014}.
The new configuration of the system $q$ depends only on the previous one, while its duration $\tau$ depends only on the new $q$. The transition function can then be written as
\begin{align}
    P(q',\tau'|q,\tau) = P^\mathrm{t}(\tau'|q') P^\mathrm{q}(q'|q)
\end{align}

\begin{figure}[H]
    \centering
    \includegraphics[width=1.0\linewidth]{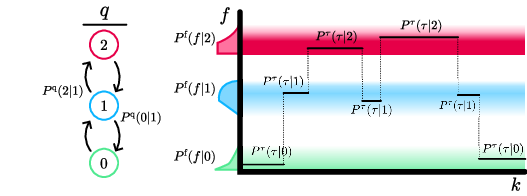}
    \caption{Evolution of the state and friction of an FSS. The left nodes show the three configurations and the transitions between them. The right side shows the friction distributions of each configuration, together with a possible trajectory of the friction value generated by the FSS as it jumps between configurations.}
    \label{fig:switching-system-diagram}
\end{figure}

The transfunctions $P^\mathrm{q}$ and $P^\tau$ can be further restricted. First, we assume an FSS can only transition to adjacent configurations, that is, $P^\mathrm{q}(q'|q) = 0$ if $|q' - q| \neq 1$. The index $q$ of the configuration is related to the friction: a higher $q$ implies a higher friction $f$, and there is no overlap in the range of values of friction produced by two different configurations.
Figure~\ref{fig:switching-system-diagram} shows an example of friction probabilities, in different colors for each $q$.

To model the behavior of an RWA, we use several FSS models, each indexed by an integer $s$. Each FSS is consequently characterized by three functions: the configuration transition function $P_s^\mathrm{q} (q|q')$ that determines how the configuration can evolve, the friction function $P_s^\mathrm{f}(f|q)$ that associates to each configuration the friction it can produce, and the duration function $P_s^\mathrm{t} (\tau|q)$ that determines how long the system stays in a given configuration.

\begin{figure}
    \centering
    \includegraphics[width=1.0\linewidth]{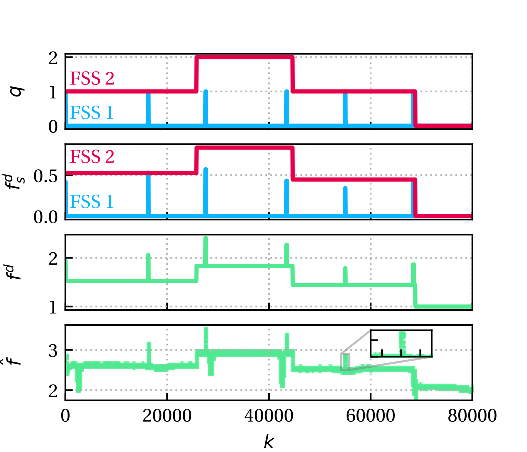}
    \caption{
        Visualization of the two friction systems described in Examples~1 (FSS 1) and 2 (FSS 2). The first two plots show the evolution of the state variables of the system. The third plot shows the resulting total dry friction as in \eqref{eq:model_dry-friction}, and the fourth the total friction after adding a viscous component and noise, as in \eqref{eq:model_friction}.
    }
    \label{fig:switching-system}
\end{figure}

An example of the friction produced by an RWA modeled with two FSS models is shown in Figure~\ref{fig:switching-system}, where its similarity to the examples shown in Figure~\ref{fig:example_ds} can be noted. 

\begin{example}[Short-term friction changes] \label{ex:hmm}
    We want to model sudden and short-lived increases of friction. This system has two configurations: inactive ($q=0$, no friction increase) and active ($q=1$, increased friction). The system stays inactive most of the time, getting activated only for a few steps before jumping back. Such behavior have been hypothesized to be caused by accumulations of oil pockets in the RWA bearings \cite{lee_-flight_2015}, and can be modeled with the functions
    \begin{subequations}
        \begin{align}
            P_1^\mathrm{f} (f | q) &\sim \mathcal{U}_\mathrm{c}[0.3q, 0.6q]
            \\
            P_1^{\mathrm{t}} (\tau| q) &\sim \begin{cases}
                \mathcal{U}_\mathrm{d}\{10000, 20000\} \quad &\text{if } q = 0
                \\
                \mathcal{U}_\mathrm{d}\{0, 200\} \quad &\text{if } q = 1
            \end{cases}
            \\
            P_1 (q'|q) &= \mathbb{1}_{q \neq q'}/\textstyle{\sum_q} \mathbb{1}_{q' \neq q}
        \end{align}
    \end{subequations}
\end{example}
where $\mathcal{U}_\mathrm{d}\{a, b\}$ and $\mathcal{U}_\mathrm{c}[a, b]$ are uniform discrete and continuous distributions, respectively, supported in the interval $[a, b]$. The symbol $\mathbb{1}_p$ represents the indicator function, equal to $1$ when $p$ is true and $0$ otherwise.

\begin{example}[Long-term friction changes]
    We want to model long-term variations of the friction switching between three configurations, with the same transition probabilities. Such variations can be due to, for example, changes in the angle of the cage inside the bearing. The probability functions are given by:
    \begin{subequations}
        \begin{align}
            P_2^{\mathrm{f}} (f | q) &\sim \mathcal{U}_\mathrm{c}[0.4q, 0.6q]
            \\
            P_2^{\tau} (\tau|q) &\sim \mathcal{U}_\mathrm{d}\{10000, 30000\}
            \\
            P_2 (q'|q) &= \mathbb{1}_{q \neq q'}/\textstyle{\sum_q} \mathbb{1}_{q' \neq q}
        \end{align}
    \end{subequations}
\end{example}

Consider the friction system proposed in Equations~\eqref{eq:model_friction} and \eqref{eq:model_dry-friction} with the FSS defined by the probability functions described above with parameters $f^{\bar{\mathrm{d}}} = f^\mathrm{v} = 1$. The evolution of the friction components, along with the total dry friction coefficient, is shown in Figure~\ref{fig:switching-system}.

%


\begin{remark}\label{rem:we_know_p}
    In the paper, we assume knowledge of $q_{\text{max}}$, $P_s^\tau (\tau|q)$ and $P_s^{\mathrm{q}} (q| q')$. Such information may be derived from historical data, expert knowledge, or both.
    We do not assume that the friction probability function $P_s^{\mathrm{f}} (f|q)$ is known.
\end{remark}

\subsection{Anomaly Modeling}\label{sec:anomaly-model}

To conclude this section, we define how anomalies in the RWA are represented in the proposed model. As mentioned in Section~\ref{sec:introduction}, several possible anomalies are considered, and all of them affect the friction experienced in the bearing of the RWA.

We consider each anomaly to relate to one friction component. For instance, the dry friction anomaly $\theta^\mathrm{d}$ affects the base dry friction coefficient, while the viscous friction anomaly $\theta^\mathrm{v}$ affects the viscous friction coefficient.

Each FSS can be affected by one anomaly, represented by $\theta_s$, where $s$ is the index of the affected FSS. We assume that the effect of the anomaly manifests itself only in the friction distribution function, that is, in the values of the friction the system produces for every configuration. This relationship can be expressed as:
\begin{subequations}
    \label{eq:anomaly_model}
    \begin{align}
        f^{\bar{\mathrm{d}}} = f^{\bar{\mathrm{d}}} (\theta^\mathrm{d}),
        \quad
        f^\mathrm{v} = f^\mathrm{v} (\theta^\mathrm{v}),
        \\
        P_s^{\mathrm{f}} (f^* | x^*) = P_s^{\mathrm{f}} (f^* | x^*; \theta^s).
    \end{align}
\end{subequations}
Although the probability functions are unknown, they can be estimated based on the dataset $\mathcal{D}$.


\section{Anomaly Detection Algorithm}\label{sec:methods}
The proposed anomaly detection algorithm uses the model developed in Section~\ref{sec:model} to analyze the window of measurements under investigation and diagnoses the status of the RWA. The possible anomalies affect the friction values that each component (base dry, viscous and the FSSs) produces. The objective, then, is to extract the relevant information from the data to do the classification.



The anomaly detection algorithm comprises four stages. The first stage, \emph{changepoint detection}, determines the time-steps in which an abrupt friction change has occurred. The second stage is the \emph{friction estimation}, which determines the dry and viscous friction coefficients before and after every changepoint, that is, $f_k^\mathrm{d}$ and $f^\mathrm{v}$ in \eqref{eq:model_friction}. The friction coefficients are used to calculate how much the friction changes at every changepoint.
The third stage is the \emph{maximum likelihood assignment}, which determines which FSS has produced each changepoint by maximizing the likelihood of the observed data. With this information, we can distinguish the individual contributions $f^{\bar{\mathrm{d}}}$ and $f_k^s$ in \eqref{eq:model_dry-friction}.
Finally, once the dry friction, viscous friction, and friction produced by the switching systems is known, the \emph{anomaly classification} stage produces an estimated status $\hat{\theta}$ using an SVM trained on the dataset provided. A diagram illustrating the stages of the algorithm, along with the data flow between them, is shown in Figure~\ref{fig:full-diagram}.

\begin{figure*}
    \centering
    \begin{tikzpicture}
    \node[simple_node] (cd) at (2.0, 0) {Changepoint\\Detection};
    
    \node[simple_node] (fe) at (6.0, 0) {Friction\\Estimation};
    
    \node[simple_node] (ia) at (10.0, 0.0) {Maximum Likelihood\\Assignment};
    
    \node[simple_node] (fc) at (14.0, 0.0) {Anomaly\\Classification};

    \path[simple_line] (0.0, 0) -- node[below, yshift=-4mm] {
        $\left\{ (\omega_k, f_k) \right\}_{k=1}^{N_\mathrm{m}}$
    } (cd);
    
    \path[simple_line] (cd) -- node[below, yshift=-4mm] {
        $\left\{k_z^\text{jump}\right\}_{z=1}^{N_\text{jump}}$
    } (fe);
    
    \path[simple_line] (fe) -- node[below, yshift=-4mm] {
        $\hat{F}, \hat{f}^\mathrm{v}, \left\{c_i^\text{skip}\right\}_{i=1}^{N_\text{cp}}$
    } (ia);
    
    \path[simple_line] (ia) -- node[below, yshift=-4mm] {
        $
        \left(
            \hat{f}^{\bar{\mathrm{d}}}
            ,
            \hat{f}^\mathrm{v}
            ,
            \left\{
                \mathcal{Q}_s,
                \mathcal{F}_s
            \right\}_{s=1}^{N_s}
        \right)
        $
    } (fc);
    
    \path[simple_line] (fc) -- node[below, yshift=-4mm] {
        $\hat{\theta}$
    } (16.0, 0.0);

    \path[simple_line] (0.5, 0) -- (0.5, 1.0) -- (6.0, 1.0) -- (fe);
    
    \path[simple_line] (4.0, 0) -- (4.0, 1.2) -- (10.0, 1.2) -- (ia);
    

     \filldraw [black]
         (0.5,0) circle [radius=1.5pt]
         (4.0,0) circle [radius=1.5pt]
     ;

\end{tikzpicture}
    \caption{Structure of the proposed anomaly detection algorithm. It consists of 4 stages. The changepoint detection determines the location of the sudden changes of friction $k_z^\mathrm{jump}$. The friction estimation estimates the friction coefficients between changepoints. The interval assignment assigns the changepoints to the different FSS in the satellite. Finally, the friction classification uses the friction values associated to the FSS to determine the anomalies present in the RWA.}
    \label{fig:full-diagram}
\end{figure*}



\subsection{Changepoint Detection}\label{sec:method_changepoints}
The first stage of the algorithm is the changepoint detection, which aims to determine at which steps $k$ in the signal there has been a sudden change in the total dry friction coefficient $f^\mathrm{d}_k$.
This step, then, takes as input the signal from the on-board data of the satellite $\{(\omega_k, \hat{f}_k)\}_{k=1}^{N_\mathrm{m}}$ and produces as output a list of points where a sudden change of friction has been detected, defined as $\left\{k^{\mathrm{cp}}_i\right\}_{i=1}^{N_\mathrm{cp}}$, where $N_\mathrm{cp}$ is the number of detected changepoints.


First, we compute a metric that estimates the probability that the friction has changed for every step $k$ in the data. Then, we find the local maxima of this metric that are above a certain threshold. These maxima are our estimates of where the changepoints are located.

Starting with the metric, we use a normalized windowed Generalized Likelihood Ratio (wGLR) \cite{gustafsson_adaptive_2000} with priors, defined as:
\begin{equation}
    \begin{array}{l}
        \displaystyle \text{wGLR}_k = \log \frac{P(X \ | \ \text{jump at } k)}{P(X |\text{no jump at } k)}, \\
        \displaystyle P(X \ | \ \text{jump at } k)
        = \max_{f^{\mathrm{d}_1}, f^{\mathrm{d}_2}, f^\mathrm{v}} \Bigl\{ P(X_{k-w:k} | f^{\mathrm{d}_1}, f^\mathrm{v}) \\
        \displaystyle \hspace*{4cm} \times P(X_{k:k+w} \ | \ f^{\mathrm{d}_2}, f^\mathrm{v}) P(f^\mathrm{v}) \Bigr\}, \\
        \displaystyle P(X \ | \ \text{no jump at } k) = \max_{f^\mathrm{d}, f^\mathrm{v}} P(X_{k-w:k+w} | f^\mathrm{d}, f^\mathrm{v}) P(f^\mathrm{v}),
    \end{array}
    \label{eq:glr}
\end{equation}
where
\begin{subequations}
    \begin{align}
        \log P (X_{a:b}\ |\ f^\mathrm{d}, f^\mathrm{v})
        &=
        \sum_{k=a}^b C \left[
            f_k - f^\mathrm{d} - f^\mathrm{v} \omega_k
        \right]^2,
        \\
        \log P(f^\mathrm{v}) &= W_\mathrm{b} \left(f^\mathrm{v} - \tilde{f}^\mathrm{v}\right)^2,
    \end{align}
\end{subequations}
with $C$ being a constant that gets canceled in the computation of the wGLR, and $W_\mathrm{b}$ a parameter used to penalize large deviations of the viscous friction from the values that we expect to find in the data. 

\begin{remark}
    Similarly to the Log-Likelihood Ratio (LLR), the wGLR metric computes the log-likelihood of two hypotheses: a jump taking place at $k$, or no jump at all. One of the issues with the LLR is the need to consider the variables involved in both hypotheses, that is, the dry and viscous friction coefficients. This problem can be addressed through marginalization, leading to the Marginalized Likelihood Ratio or through maximization for the parameters, as we do in this problem, leading to the Generalized Likelihood Ratio \cite{gustafsson_adaptive_2000}.
\end{remark}

The GLR was developed for the case when there is only one changepoint to detect, but our problem contain several. Rather than computing the likelihood for the whole dataset, this issue is addressed by computing it in a window of $w$ points around the hypothetical changepoints. The constant $w$ is a hyperparameter selected from historical data, on the condition that this window size is smaller than the minimum distance between changepoints expected in the data.

The last element in the metric is the use of a prior for the viscous friction $P(f^\mathrm{v})$. Its purpose is to avoid obtaining parameters in the optimization of \eqref{eq:glr} that fit the data well but are not realistic. As we will see in Section~\ref{sec:tuning-wb}, a high penalization constant $W_\textrm{b}$ can improve the algorithm's effectiveness when the window size $w$ is small. The base viscous friction coefficient $\bar{f}^\mathrm{v}$ is obtained from historical data and, as it will be explained also in Section~\ref{sec:tuning-wb}, does not need to be very accurate to fulfill its objective.

Once the wGLR has been computed, we must determine which points correspond to changepoints. The original GLR algorithm solves this problem by taking the point $k$ at which the metric is maximized.
In our case, however, we must take into account that there are several changepoints in the complete dataset. Therefore, we first find windows of contiguous measurements where the $\text{GLR}$ is above a threshold $\text{GLR}_\text{thr}$ and, for each window, we designate the maximum among those points to be a changepoint $k^\text{cp}$.

The effectiveness of this stage of the algorithm depends on how the threshold $\text{GLR}_\mathrm{thr}$ is set.
Under the assumption that the dry friction coefficient is constant and the viscous friction coefficient is correctly estimated with $\tilde{f}^\mathrm{v}$, Wilk's theorem shows that $\text{GLR}/\sigma_v^2 \sim \chi^2$, where $\sigma_v^2$ is the variance of the noise in the signal \cite{wilks_large-sample_1938}. This provides a good way to determine the threshold: if we want to have a false positive probability smaller than $10^{-9}$, we find the value of the threshold such that $P(\chi^2 > \text{GLR}_\text{thr}/\sigma_v^2) < 10^{-9}$. However, a high threshold will also reduce the probability of detecting changepoints, a  trade-off that will be analyzed in Section~\ref{sec:results}.


The output of the algorithm is the list $\left\{k^\text{cp}_i\right\}_{i=1}^{N_\text{cp}}$ of all changepoints detected in the window of measurements.

\subsection{Friction Estimation}\label{sec:friction}
Once the changepoints have been obtained, we determine the dry friction coefficient between changepoints, together with the viscous friction in the data. Since the friction model we are using is linear in both friction parameters, we can use least-squares to estimate these parameters.

This stage of the algorithm takes the original signal $X = \left\{\left(\omega_k, f_k\right) \right\}_{k=1}^{N}$, together with the changepoint locations estimated in the previous step $\left\{k_i^\text{cp}\right\}_{i=1}^{N_\text{cp}}$. The algorithm's output is a vector with the value of the dry friction coefficient at each interval, together with the viscous friction, which was assumed to remain constant along the whole trajectory being analyzed.

First, we determine the regions in which the dry friction coefficient stays constant, that is, the intervals between changepoints. Each interval $i$ is defined by the first ($k_i^o$) and last ($k_i^f$) measurement included in it. These measurement indexes are computed as:
\begin{align}
    k_i^\mathrm{o} &= \max\left(k_{i-1}^\mathrm{cp}, 0\right)
    \\
    k_i^\mathrm{f} &= \min\left(k_{i}^\mathrm{cp}, N\right)
\end{align}

Due to the noise affecting the measured signals, a changepoint may be detected some steps before or after the actual change in the friction coefficient. We ignore the points near the changepoints to avoid the negative effects that these may have on estimating these coefficients. We can take $\Delta k_{\mathrm{error}} = w/2$ for small window sizes, which has been observed to give good results. With this process we obtain a total of $N_{\mathrm{int}}$ intervals, where $N_{\mathrm{int}} = N_{\mathrm{cp}} + 1$.

For each interval $i$, the model in \eqref{eq:model_friction} can be written as
\begin{align}
    \label{eq:interval_friction}
    \hat{f}_k &= f_i^\mathrm{d} + f_v \omega_k + v.
\end{align}
We collect all the dry friction coefficients $f_i^\mathrm{d}$ in a vector $F$ and define the observation matrix $H_i$ and estimated friction vector $Y_i$ for each interval as follows:
\begin{align}
    Y_i &= \begin{bmatrix}
        \hat{f}_{k_i^o} & \hat{f}_{k_i^o + 1} & \cdots & \hat{f}_{k_i^f}
    \end{bmatrix}^T,
    \\
    H_i &=
    \left[
    \begin{array}{ccc|c}
    & & & \omega_{k_i^\mathrm{o}}\\
    \mathbb{0}_{\Delta k_i \times i} & \mathbb{1}_{\Delta k_i \times 1} & \mathbb{0}_{\Delta k_i \times (N_\text{int}-i)} & \vdots \\
    & & & \omega_{k_i^\mathrm{f}} \\
    \end{array}
    \right],
    \\
    F &= \begin{bmatrix}
        f_1^d & f_2^d & \cdots  & f_{N_\mathrm{int}}^\mathrm{d}
    \end{bmatrix}^T,
\end{align}
where $\Delta k_i = k_i^\mathrm{o} -k_i^\mathrm{f} + 1$ is the number of points in interval $i$, and $\mathbb{0}_{a \times b}$ and $\mathbb{1}_{a \times b}$ are matrices of shape $a \times b$ filled with zeros and ones, respectively.
Then, \eqref{eq:interval_friction} can be written in matrix form:
\begin{align}
    Y_i &= H_i \begin{bmatrix} F^T & f_v\end{bmatrix} + v.
\end{align}

We can then concatenate the observation matrices and estimated friction vectors to produce a single equation for all measurements:
\begin{align}
    H &= \begin{bmatrix}
        H_1 \\ H_2 \\ \vdots \\ H_{N_\mathrm{int}}
    \end{bmatrix}
    \quad
    &Y &= \begin{bmatrix}
        Y_1 \\ Y_2 \\ \vdots \\ Y_{N_\mathrm{int}}
    \end{bmatrix}
\end{align}
to give
\begin{align}
    \label{eq:algebraic_ls}
    Y = H \begin{bmatrix} F^T & f^\mathrm{v} \end{bmatrix}^T + v
\end{align}

Since Equation~\eqref{eq:algebraic_ls} is linear in $F$ and $f^\mathrm{v}$, we use least-squares to obtain an unbiased estimate of $F$ and $f^\mathrm{v}$:
\begin{equation*}
    \begin{array}{rl}
        \begin{bmatrix} \hat{F}^T & \hat{f}^\mathrm{v} \end{bmatrix}
    &= Y^T H \left(H^T H\right)^{-1}.
    \end{array}
\end{equation*}

There is a probability that the changepoint detector is triggered at false positives. To analyze the likelihood that a certain changepoint was incorrectly detected, we define the metric $c_i^\text{rjct}$, corresponding to the increase in the squared error of the friction estimation if the changepoint $i$ is rejected (that is, ignored), and intervals $i$ and $i+1$ are assumed to have the same friction coefficient. It is computed as:
\begin{align}
    c_i^\text{rjct} = &\frac{
        (H^T H)_{i:i} (H^T H)_{(i+1):(i+1)}
    }{
        (H^T H)_{i:i} + (H^T H)_{(i+1):(i+1)}
    } (\hat{F}_i - \hat{F}_{i+1})^2 \nonumber
    \\ - &\log P(\mathrm{false\ positive}),
    \label{eq:rejection_cost}
\end{align}
where the probability of a false positive can be obtained from threshold $\mathrm{GLR}_\mathrm{thr}$ used for the changepoint detection, as explained at the end of Section~\ref{sec:method_changepoints}. The rejection cost will be used in the next step to decide which changepoints to consider as false positives.

The output of this stage is then the estimated dry friction at each interval, the estimated viscous friction, and the rejection cost of every changepoint: $(\hat{F}, \hat{f}^\mathrm{v}, \{ c_i^\mathrm{rjct}\}_{i=1}^{N_\mathrm{cp}})$.

With the total dry and viscous friction torques at each interval obtained, the next stage will focus on determining the individual contribution of each FSS to the friction.


\subsection{Maximum Likelihood Assignment}
Once the total dry friction coefficient $\hat{f}^\mathrm{d}_i$ has been estimated for each interval, we are interested in decomposing this friction among the FSSs, that is, to obtain the value of the various components of \eqref{eq:model_dry-friction} at every step, rather than only their sum. To achieve this, we use the knowledge of the transition probability of the subsystems $P_s^\mathrm{q}(q|q')$ and $P_s^\tau (\tau|q)$, together with the steps $k_i^\mathrm{cp}$ at which the friction changes took place to determine the most likely sequence of configurations that the subsystems went through.

Once this sequence has been estimated, we determine the values of each subsystem's friction coefficients $\hat{f}_k^s$ and the base dry friction $\hat{f}^{\bar{\mathrm{d}}}$. Since the dry and viscous coefficients are a function of the anomalies present in the RWA, as in \eqref{eq:anomaly_model}, we use their values for the diagnosis.

Formally, the algorithm takes the friction values, rejection costs, and location of the changepoints as input. It estimates the base dry friction, and the friction values of each subsystem at each configuration (that is, all the values in \eqref{eq:model_dry-friction}).

\subsubsection{Maximum likelihood jumping}

The first step is to determine which FSS has produced each one of the changepoints detected by the first stage of the algorithm. The output, then, is a vector $u \in \mathbb{N}^{N_\mathrm{cp}}$ where each entry $u_i$ is the index of the FSS that has caused the changepoint $i$. The objective is to determine the $u$ such that the jumps have the highest likelihood, with the probabilities given by the functions $P_s^\mathrm{t}$ and $P_s^\mathrm{q}$.

The log-likelihood of a specific sequence of jumps can be computed sequentially, which makes dynamic programming a natural choice to find the vector $u$ that maximizes it. 
A multistage dynamic programming is defined by four elements: a state set ($\mathcal{X}$), an input set ($\mathcal{U}$), a transition function (\hbox{$h: \mathcal{X} \times \mathcal{U} \mapsto \mathcal{X}$}), and a cost function ($g: \mathcal{X} \times \mathcal{U} \mapsto \mathbb{R}$) \cite{bertsekas_dynamic_2012}. The objective is to solve the following optimization problem:
\begin{subequations}
    \begin{align}
        u^* =& \argmax_u \sum_{i=1}^{N_c} g(x_i, u_i)
        \\
        \text{s.t.}&\quad x_{i+1} = h_i(x_i, u_i)
    \end{align}
    \label{eq:dp_problem}
\end{subequations}

Extending the notation used in Section~\ref{sec:fss}, we define the state of the RWA at every interval $i$ with a tuple of two elements: the configuration $\bar{q}_i$ of each subsystem and the number of steps $\bar{\tau}_i$ since each subsystem changed configuration for the last time, both given as vectors with as many elements as subsystems present in the RWA, so $x_i = (\bar{q}_i, \bar{\tau}_i)$, with
$\bar{q}_i = \begin{bmatrix} q_i^1 & \cdots & q_i^{N_\mathrm{s}}\end{bmatrix}^T$
and
$\bar{\tau}_i = \begin{bmatrix} \tau_i^1 & \cdots & \tau_i^{N_\mathrm{s}} \end{bmatrix}^T$.

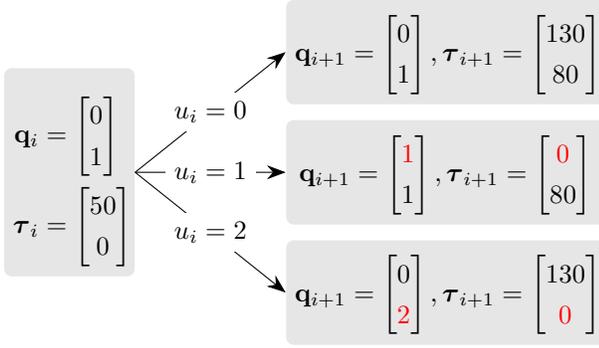
\begin{figure}[ht]
    \centering
    \begin{tikzpicture}

    \node[text width=15mm, anchor=center, fill=black!10, rounded corners=1mm] (start) at (-2, 1.8) {
        \begin{align*}
            \mathbf{q}_i &= \begin{bmatrix}
                0 \\ 1
            \end{bmatrix}
            \\
            \bar{\tau}_i &= \begin{bmatrix}
                50 \\ 0
            \end{bmatrix}
        \end{align*}
    };

    \node[text width=40mm, anchor=center, fill=black!10, rounded corners=1mm] (a) at (3, 1.8) {
        \vskip-6mm
        \begin{equation*}
            \bar{q}_{i+1} = \begin{bmatrix}
                \textcolor{red}{1} \\ 1
            \end{bmatrix}
            ,
            \bar{\tau}_{i+1} = \begin{bmatrix}
                \textcolor{red}{0} \\ 80
            \end{bmatrix}
        \end{equation*}
    };
    
    \node[text width=40mm, anchor=center, fill=black!10, rounded corners=1mm] (b) at (3, 0.2) {
        \vskip-6mm
        \begin{equation*}
            \bar{q}_{i+1} = \begin{bmatrix}
                0 \\ \textcolor{red}{2}
            \end{bmatrix}
            ,
            \bar{\tau}_{i+1} = \begin{bmatrix}
                130 \\ \textcolor{red}{0}
            \end{bmatrix}
        \end{equation*}
    };
    
    \node[text width=40mm, align=center, fill=black!10, rounded corners=1mm] (d) at (3, 3.4) {
        \vskip-6mm
        \begin{equation*}
            \bar{q}_{i+1} = \begin{bmatrix}
                0 \\ 1
            \end{bmatrix}
            ,
            \bar{\tau}_{i+1} = \begin{bmatrix}
                130 \\ 80
            \end{bmatrix}
        \end{equation*}
    };


    \path[simple_line] (start.east) -- node[fill=white] {$u_i=1$} (a.west);
    \path[simple_line] (start.east) -- node[fill=white] {$u_i=2$} (b.west);
    \path[simple_line] (start.east) -- node[fill=white] {$u_i=0$} (d.west);


    
\end{tikzpicture}
    \caption{Example of possible transitions during the dynamic programming problem, given that the next changepoint has $\Delta f_i = +1$, and $\Delta \tau_i = 80$.}
    \label{fig:transition-diagram}
\end{figure}

With this notation, $q_i^s$ represents the configuration of the FSS with index $s$ at interval $i$.
Every changepoint $i$ is characterized by two variables: the number of steps since the previous changepoint ($\Delta \tau_i$), and the sign of the friction change ($\Delta f_i$), which can be either $-1$ for a decrease or $1$ for an increase of the friction. At a changepoint, one of the subsystems changes its configuration by $\Delta f_i$ and its number of steps since the last changepoint is reset to $0$. The other subsystems have unaffected configurations, but their $\tau$ increases by $\Delta \tau_i$.
We choose which subsystem has jumped at each stage with the input $u_i$. The set of possible inputs is $\mathcal{U} \in [1, N_\textrm{s}]$, and the transition function is, according to the description given before,
\begin{align}
    h_i (\bar{x}_i, u_i)
    &=
    \left(
        \bar{q}_i + \Delta f_i \bar{e}_{u_i},
        (\bar{\tau}_i + \Delta \tau_i) (\mathbb{1} - \bar{e}_{u_i})
    \right),
\end{align}
where $\mathbb{1}$ is a vector with all entries equal to $1$ and $\bar{e}_u$ is a vector with all entries $0$ except the $u$-th one, equal to $1$. A visualization of these transitions is shown in Figure~\ref{fig:transition-diagram}.

Regarding the transition costs of the problem, they are given by the log-likelihood of the transitions of the individual FSS:
\begin{align}
    g_i(x_i, u_i) &= \sum_{s\not=u_i} \log P_s^{\tau} (\tau \geq \tau_s+\Delta \tau|q, \tau \geq \tau_s) \nonumber\\ 
    + &\log P_{u_i}^{\tau} (\tau = \tau_{u_i} + \Delta \tau_i |q_{u_i}, \tau \geq \tau_{u_i}) \label{eq:dp_score}\\ 
    + &\log P_{u_i}^{\mathrm{q}}(q_{u_i} + \Delta f_i|q_{u_i}).
    \nonumber
\end{align}
In this computation, conditional probabilities of $\tau$ are computed as $P_s^\tau(\tau \geq \tau_1 | q, \tau_2) = P(\tau \geq \tau_1|q)/P(\tau \geq \tau_2|q)$ when $\tau_1 \geq \tau_2$. In a similar way, $P_s^\tau(\tau = \tau_1 | q, \tau_2) = P(\tau = \tau_1|q)/P(\tau \geq \tau_2|q)$ under the same conditions.

Recalling Remark~\ref{rem:we_know_p}, we know $P_s^\tau(q, \tau)$ and $P_s^\mathrm{q}(q'|q)$, so we can compute all the log-likelihood scores, either as the dynamic programming problem is being solved, or in advance, and then use look-up tables during the execution of the algorithm.
The objective, then, is to obtain an input sequence $u$ that maximizes the log-likelihood. This is a standard dynamic programming problem that can be solved with classical algorithms \cite{bertsekas_dynamic_2012}. Note that while the possible values of $\bar{q}$ are relatively limited, this is not the case for $\bar{\tau}$, and the number of possible states may increase exponentially with the number of stages.

For the initial state $x_0$ of the dynamic programming problem, we take an optimistic approach: for $\bar{q}_0$, we initialize the algorithm considering all possible values, and keep the one that provides the maximum final score. Regarding $\bar{\tau}_0$, we select the value that maximizes the score in \eqref{eq:dp_score}.

The solution $u^*$, obtained by solving \eqref{eq:dp_problem}, provides then the index of the FSS that produced each sudden change of friction observed in the data. Additionally, we denote with $\bar{x}^*$ the sequence of states obtained by applying this input.

\subsubsection{Changepoint rejection}\label{sec:rejection}

The algorithm presented is inherently fragile: if a changepoint is missed during the first stage of the algorithm, or a false positive is obtained, the algorithm will fail to obtain the correct sequence of configurations of the FSSs. 
Since we want the algorithm to perform well even in the presence of false positives, we robustify it by including in the dynamic programming problem the option at every stage of ignoring the changepoint detected, rather than fitting it in one of the FSSs. This is done by including $0$ as an additional possible input, shown in Figure~\ref{fig:transition-diagram}, with the following transition and cost functions:
\begin{align}
    h_i(x_i, 0) 
    &=
    \left(
        q_i, \tau_i + \mathbb{1} \Delta \tau_i
    \right).
\end{align}
The cost of this transition is given by the log-likelihood of none of the switching systems jumping during this stage, plus the added cost of rejecting this changepoint. Formally, this is written as
\begin{align}
    g(x_i, 0) &= 
    c_i^\text{rjct} + \sum_{s} P^{\tau}(\tau \geq \tau_s + \Delta \tau_i|q_s, \tau \geq \tau_s).
\end{align}

Recalling \eqref{eq:rejection_cost}, this additional cost is given by how much the squared error of the regression increases when the changepoint is removed and the same friction coefficient is estimated before and after it. Therefore, the algorithm is more likely to ignore changepoints when the intervals before and after them have similar friction coefficients.

\subsubsection{Friction Values}

Once the dynamic programming algorithm is finished, we get a vector $u$ containing the index of the FSS that changed configuration at each changepoint and an array $\bar{q}$ with the configuration of every FSS after every changepoint. The next objective is to estimate the base dry friction $\hat{f}^{\bar{\mathrm{d}}}$ and, for each FSS, the sequence of configurations and friction values it goes through. We denote them as $\mathcal{Q}_s$ and $\mathcal{F}_s$ respectively, both defined as lists of values in the programming sense, where $s$ is the index of the FSS. At every interval, the total dry friction torque must be the base dry friction plus the contribution of each FSS as given by \eqref{eq:model_dry-friction}.

This problem is solved in Algorithm~\ref{alg:friction_values}. We first assume that, before the first changepoint, the friction produced by each FSS is equal to $0$. Then, we compute the following friction values from the differences obtained at each changepoint. Finally, since the minimum value produced by an FSS is assumed to be $0$, we adjust the coefficients to get the final values.

\begin{algorithm}
    \begin{algorithmic}[1]
        \Require $\hat{f}_i$, $u_i$, $q_i^s$
        \LineComment{Assume that each FSS starts with $f^s = 0$}
        \State $\mathcal{F}_s = [0] \quad \forall s \in [1, N_s]$
        \State $\mathcal{Q}_s = [q_1^s] \quad \forall s \in [1, N_s]$
        \LineComment{At start, all friction is assumed to be the base dry friction}
        \State $\hat{f}^{\bar{d}} = \hat{f}_1$
        \For{$i \gets 2$, $N_\text{int}$}
            \LineComment{At every changepoints, add the $f$ and $q$ to the corresponding FSS.}
            \State $\mathcal{F}_{u_i}$.append($\mathcal{F}_{u_i}[-1] + \hat{f}_i - \hat{f}_{i-1}$)
            \State $\mathcal{Q}_{u_i}$.append($q_i$)
        \EndFor
        \For{$s \gets 1$, $N_s$}
            \LineComment{Adjust $\mathcal{F}_s$ so its minimum is equal to $0$}
            \State $f_{\text{min}} = \min \mathcal{F}[s]$
            \For{$i \gets 1$, $\text{len}(\mathcal{F}[s])$}
                \State $\mathcal{F}_s[i] \gets \mathcal{F}_s[i] - f_{\text{min}}$
            \EndFor
            \State $f_\text{base} \gets f_s + f_{\text{min}}$
        \EndFor
    \end{algorithmic}
    \caption{Obtain $\mathcal{Q}_s$ and $\mathcal{F}_s$}
    \label{alg:friction_values}
\end{algorithm}


The output of this stage is a tuple of four elements: the estimated base dry friction ($\hat{f}^{\bar{d}}$, the estimated viscous friction coefficient ($\hat{f}^v$), and the sequence of states ($\mathcal{Q}_s$) and friction values ($\mathcal{F}_s$) each FSS goes through: 
$
    (
        \hat{f}^{\bar{\mathrm{d}}}
        ,
        \hat{f}^\mathrm{v}
        ,
        \{
            \mathcal{Q}_s, \mathcal{F}_s
        \}_{s=1}^{N_\mathrm{s}}
    )
$.

\subsection{Anomaly Classification}
The final stage is the anomaly classification, in which the friction coefficients obtained in the last step are used to determine the anomaly status of each component affecting the friction torque in the RWA.

This step, then, takes the assigned friction coefficients $\hat{f}^{\bar{d}}$, $\hat{f}^v$ and $\left\{(\mathcal{Q}_s, \mathcal{F}_s)\right\}$ for each FSS $s$, and produces a diagnosis for the dry friction $\hat{\theta}_d$, a diagnosis for the viscous friction $\hat{\theta}_v$ and one diagnosis for each switching system $\hat{\theta}_s$. These diagnoses are independent from each other: $\hat{\theta}_d$ depends only on $\hat{f}^{\bar{d}}$, $\hat{\theta}_v$ on $\hat{f}^v$, and the diagnosis of each subsystem is done with the friction values associated to it. We can then treat each of these classification problems as independent.

As mentioned in Section~\ref{sec:model}, we have knowledge of which part of the friction model is affected by each anomaly; however, we do not know how it is affected. To address this shortcoming, we use the labeled dataset $D$, where we have a collection of datapoints with the anomalies associated with them. For each datapoint in the dataset, we run the first three steps of the algorithm to obtain a processed dataset
\begin{align}
    D_\mathrm{p}
    &= 
    \left\{\left(
        \hat{f}^{\bar{\mathrm{d}}},
        \hat{f}^\mathrm{v},
        \left\{
            \left(
                \mathcal{Q}_s, \mathcal{F}_s
            \right)
        \right\}_{s=1}^{N_\mathrm{s}}
    \right)\right\}_{i=1}^{N_\mathrm{D}}.
\end{align}
This dataset we can be split into $2 + N_s$ different datasets, one for each anomaly:
\begin{subequations}
    \begin{align}
        D^\mathrm{\bar{d}} &= \left\{\left(
            \hat{f}^{\bar{\mathrm{d}}}, \theta^\mathrm{d}
        \right)\right\}, \qquad 
        D^\mathrm{v} = \left\{\left(
            \hat{f}^\mathrm{v}, \theta^\mathrm{v}
        \right)\right\},
        \\
        D^s &= \left\{\left(
            \left\{\left(\mathcal{Q}_s, \mathcal{F}_s \right)\right\}_{s=1}^{N_\mathrm{s}}, \theta^s
        \right)\right\} \forall s \in [1, N_s].
    \end{align}
\end{subequations}

In our problem, the status of each anomaly is given by a simple binary label: active or inactive. In our pursue of a simple and effective tool for the classification, we choose to use Support Vector Machines (SVM) \cite{hastie_elements_2009}. By defining one SVM per anomaly, we can train them on the corresponding independent dataset.

An SVM is a classification method that takes as input a vector $z \in \mathbb{R}^n$ and produces an output $y \in \{-1, 1\}$ according to the following equation:
\begin{align}
    \hat{\theta}(z) = \text{sign}\left( w^T z + b\right).
\end{align}
The classifier is parametrized by the weights $w$ and the bias $b$. Although not as versatile as state-of-the-art neural networks, SVMs have two important advantages due to their reduced number of parameters: they are less prone to overfitting, and doing inference is computationally less expensive. Although we use SVMs to classify all anomalies, the implementations vary depending on whether we are determining a dry/viscous friction anomaly, or an anomaly related to an FSS.

\subsubsection{Dry and viscous friction anomalies}
The first two anomalies involve an increased value in the base dry and viscous friction coefficients, respectively. The classification, therefore, consists in analyzing a single value in each case, either $z = \hat{f}^{\bar{\mathrm{d}}}$ or $z = \hat{f}^\mathrm{v}$.
This is accomplished by defining an SVM that takes values in $\mathbb{R}^1$ as inputs, making the classification a simple threshold learned from the dataset.

\subsubsection{Subsystem anomalies}

Classifying anomalies for switching systems is more challenging, since we do not have a single value to classify with, but a sequence of them in the form of $\mathcal{F}_s$.
As mentioned in Subsection~\ref{sec:anomaly-model}, an anomaly impacts the probability distribution of the friction produced by the FSS, so creating a histogram of the values obtained and using an SVM that does the classification is a reasonable approach.

The normalized histogram vector of the FSS with index $s$ is defined as $z \in \mathbb{R}^{n_\text{bins}}$, obtained from a sequence of friction values $\mathcal{F}_s$ as
\begin{subequations}
    \begin{align}
        \forall j \in \{1, \dots, n_\text{bins}\}, \quad z_j = \frac{1}{|\mathcal{F}_s|} \sum_{f \in \mathcal{F}_s} \mathbb{1}_{f \in [r_j, r_{j+1}]},
        \\
        r_j = r_\text{min} + \frac{j}{n_\text{bins}} (r_\text{max} - r_\text{min}).
    \end{align}
\end{subequations}

Two hyperparameters must be defined, the number of bins ($n_{\text{bins}}$) and the limit values ($r_{\text{min}}$ and $r_{\text{max}}$). The latter can be determined from the historical data, by annotating the minimum and maximum values of friction observed. The number of bins to select is a trade-off between accuracy and complexity. A small number of bins may make distinguishing an anomaly impossible, while a large number will lead to a large histogram vector, which may complicate training and lead to overfitting. For our task, we select a value of $20$ bins, and later analyze the effect of this hyperparameter on the algorithm's accuracy in Section~\ref{sec:classification_results}.

The last decision to take is what range of friction values to include in the histogram. From the previous section, we recall that we obtain, for each subsystem, a collection of friction values paired with their corresponding configurations. Based on our knowledge of the anomaly, we can decide to include all friction values produced by the FSS in the histogram, or only those related to specific configurations (for example, only friction values obtained with $q=1$).

The output of this final stage is the status of the RWA that generated $X$, given as $\hat{\theta} = (\hat{\theta}^\mathrm{d}, \hat{\theta}^\mathrm{v}, \{ \hat{\theta}^s\}_{s=1}^{N_\mathrm{s}})$.

\section{Algorithm Tuning}\label{sec:tuning}
Despite the complexity of the anomaly detection algorithm, most of its parameters are fairly straight--forward to tune, or have little impact on its results. There are, however, two parameters that require special care. The first is the bias weight ($W_b$), used in the changepoint detection stage to include the effect of the prior knowledge of the viscous friction coefficient. The second is the number of bins ($n_\mathrm{bins}$) that the SVM uses to classify anomalies related to FSS. In this section, we will study the effect of these two parameteres on the performance of the algorithm, and determine the best way to tune them.

\subsubsection{Bias weight selection}\label{sec:tuning-wb}
Before deciding how to tune the bias weight, it is necessary to define what results we want to optimize.
In the analysis of a changepoint detection algorithms, there are two main metrics of performance: the Missed Detection Rate (MDR), defined as the probability of not detecting a changepoint, and the Average Run Length (ARL), defined as the average number of non-anomalous points the algorithm goes through before detecting a false changepoint in the data \cite{gustafsson_adaptive_2000}. Since it is always possible to improve one at the expense of the other (by decreasing or increasing the threshold $\text{GLR}_\text{thr}$), the challenge lies in achieving a high ARL with a low MDR.

To assess the performance of the changepoint detection, it is important to remember how the subsequent stages of the algorithm process changepoints. As mentioned in Section~\ref {sec:rejection}, an undetected changepoint can invalidate the algorithm's output while a false positive can be rejected during the assignment stage. This implies that while high ARL is beneficial, as it will lead to fewer false positives, a low MDR is required for the algorithm's success. 

For our analysis, we will fix the MDR to $0.1\%$ and determine the threshold value at which such sensitivity is achieved using the $\chi^2$ distribution. Then, we obtain the ARL with this threshold, which we call the $\text{ARL}^{0.1\%}$. A satisfactory value of $\text{ARL}$ depends on the tolerance of the algorithm to false positives, and the number of points we process in a single window of measurements. In the TAS case, we analyze windows of around $80000$ measurements, so an \text{ARL} of around $10^6$ is a good target if we do not want to get more than $1$ false changepoint per window. Since the TAS dataset has no annotated changepoint locations, these results are obtained using our own simulations, in which we can place changepoints in a controlled way.

\begin{figure}[ht]
    \centering
    \includegraphics[width=1.0\linewidth]{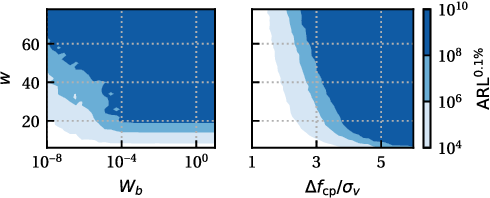}
    \caption{Left: value of $\text{ARL}^{0.1\%}$ for different window sizes $w$ and bias $W_\mathrm{b}$, assuming changes in friction of $\Delta f_\text{cp} = 3\sigma_v$. Right: effect of window size and friction change on $\text{ARL}^{0.1\%}$, with $W_b=10^{-4}$.}
    \label{fig:changepoint_plots}
\end{figure}

Back to the tuning of $W_\mathrm{b}$, we can run a parametric study of this coefficient to observe how it affects the $\text{ARL}^{0.1\%}$ for different window sizes, assuming that the changes of dry friction we are trying to detect have a magnitude of $3$ times the noise standard deviation. The results are shown in Figure~\ref{fig:changepoint_plots}, left plot.
A higher window size $w$ leads to better algorithm results, as the effect of the noise is reduced when we use more points the w-GLR. However, increasing $w$ is not always feasible, because it is limited by the minimum distance we expect to see between changepoints.
We can observe how the $W_\mathrm{b}$ improves the algorithm's performance for different window sizes. As an example, using a window size of $20$, we can improve the $ARL^{0.01\%}$ from $10^4$ to $10^{10}$ by increasing $W_\mathrm{b}$ from $10^{-8}$ to $10^{-4}$. The bias term is, then, a good way of improving the performance of the changepoint detection when the window size is constrained, and it seems from this first analysis that $W_\mathrm{b}$ should be set to be arbitrarily large.

However, in the previous plot, we assumed that our estimated friction $\tilde{f}^\mathrm{v}$ is equal to the actual viscous friction $f^\mathrm{v}$. Unfortunately, this may not always be the case, since the latter is not known in advance. 
Considering the effect of the bias $\tilde{f}^\mathrm{v} - f^\mathrm{v}$ is more complex: when the error is not zero, the $\text{w-GLR}$ no longer follows a $\chi^2$ distribution. We can, however, estimate the effect of this error to be in the order of $W_\mathrm{b} (\tilde{f}^\mathrm{v} - f^\mathrm{v})^2$. As the original distribution of the $\text{w-GLR}$ is of order unit, we should ensure that $W_\mathrm{b} (\tilde{f}^\mathrm{v} - f^\mathrm{v})^2 \lll 1$ so the error does not have a noticeable effect. Using $W_\mathrm{b} = 10^{-4}$, this implies that we need $|\tilde{f}^\mathrm{v} - f^\mathrm{v}| \lll 10^2$, an easily achievable condition since the viscous friction coefficient is of order $1$.



This gives a simple rule for tuning $W_b$: if the window size is large enough (more than 50 points), it can be set to $0$. If not, a high value can be set to improve the results, but keeping in mind that if the prior of the viscous friction is not reliable, the performance may degrade and produce too many false positives. For our implementation of the algorithm, we will use $W_b = 10^{-4}$.

\subsubsection{Number of bins}\label{sec:classification_results}

\begin{figure}[ht]
    \centering
    \includegraphics[width=\linewidth]{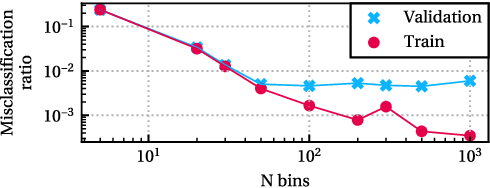}
    \caption{Train and validation loss with varying number of bins for classification of anomalies of subsystem 1.}
    \label{fig:bin-benchmark}
\end{figure}

The number of bins determines how the friction coefficients estimated for each FSS are translated into a vector that can be fed into a SVM to classify an anomaly. A small number of bins is desirable as it reduce the number of dimensions the SVM has to work with. However, it may also lead to loss of granularity in the transition from friction values to histograms, which can make the classification perform worse. As with the bias weight, we run a parametric test using different number of bins for classifying the anomaly in one of the FSS of the TAS dataset. The obtained misclassification rates for training and validation datasets is shown in Figure~\ref{fig:bin-benchmark}.

The plot shows an expected decrease in the train loss as the number of bins increases, since the dimension of the classification increases with it. However, many bins lead to overfitting, as training captures only the specific shape of the datapoints used in training, and the validation loss increases. This shows a good level of robustness of the algorithm to the number of bins, which can again be attributed to the use of SVMs for the classification. Since we see that the validation misclassification ration plateaus after around $40$ bins, we used this number in our algorithm.

\section{Experimental Results}\label{sec:results}



We now proceed to analyze the performance of the anomaly detection algorithm developed in this paper. 
The algorithm's structure, composed of four sequential stages, is beneficial for analysis, as each part of the algorithm can be analyzed independently to ensure it works as expected. This way, our analysis is not limited to the final accuracy obtained by the last stage. 

This section is organized by stages of the algorithm. We start by analyzing the changepoint detection stage using synthetic data. For the friction estimation and changepoint assignment, we use the TAS dataset to analyze some performance metrics so that, even without a ground truth to compare to, we can study the behavior of these stages. Finally, we will use the labels provided in the TAS dataset to assess the accuracy of the complete algorithm at determining the anomaly present in the RWA. Additionally, we investigate the computational cost of the whole process to establish its suitability for running in a low-power device located in the satellite.


\subsection{Changepoint detection}

We analyze the first part of the algorithm, which takes a time-series $\{(\omega_k, f_k)\}_{k=1}^{N_\mathrm{m}}$ and determines the position of the multiple abrupt changes of friction.

The capability of the changepoint detector to locate actual changepoints without being triggered with false positives depends, given a fixed $W_b$, on elements: the window size $w$ and the change of friction $\Delta f_\mathrm{cp}$ at the changepoints. We use the same metric defined in Section~\ref{sec:tuning-wb} to perform this analysis, studying the ARL obtained when the threshold of the detector is set to achieve a MDR of $0.1\%$, which we call the $\text{ARL}^{0.1\%}$. As with the tuning, we generate data artificially so we have control over where the actual changepoints are located.

The results are shown in Figure~\ref{fig:changepoint_plots}, right plot. We can see that better results are achieved with either high window sizes or changepoint strengths. This shows a basic limitation of the algorithm: if we want to detect small friction changes, we need to use large window sizes. However, this might not be possible if the changepoints that we want to detect may be too close together, in which case we need to increase the value of $W_\mathrm{b}$.

As a last note, the average run lengths obtained in this analysis assume that the data contains no abnormalities, that is, the friction is just a combination of dry and viscous friction and Gaussian noise as given by \eqref{eq:model_friction}. Real data may contain imperfections that trigger additional, unavoidable false changepoints. Therefore, a high ARL is necessary for reducing the number of false changepoints, but it does not guarantee it.

\subsection{Friction Estimation}
We now analyze the output of the second step of the algorithm: the friction estimation. In particular, we will analyze how well the friction coefficients $\hat{f}^\mathrm{d}_i$ and $\hat{f}^\mathrm{v}$ obtained in this stage approximate the friction $\hat{f}_k$ in the data.

A small error is an indicator of a good fit of the model to the data (at least when it comes to \eqref{eq:model_friction}, since we are not considering the different FSS yet) and a correct estimation of the changepoint locations. This is because, if changes in the friction were not detected, the algorithm would try to fit the same coefficient to intervals of measurements with different dry friction coefficients, which would produce a high error.

\begin{figure}[ht]
    \centering
    \includegraphics[width=\linewidth]{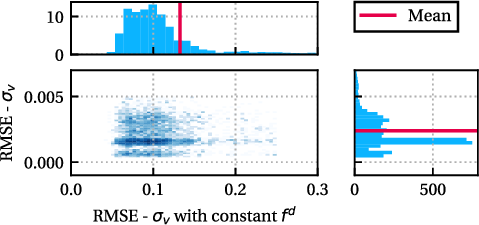}
    \caption{Excess RMSE in the datapoints of the TAS dataset. The top histogram is obtained by computing a single dry and viscous friction coefficient for the complete time-series, the right one by obtaining several dry friction coefficients from steps 1 and 2 of the algorithm. The mean values of the histograms are $0.13$ and $0.0025$, respectively.}
    \label{fig:rmse_compare}
\end{figure}

We use the Root Mean Square Error (RMSE) as a metric of fit, which measures the error between the friction values in the data and those estimated using the intervals and the linear regression run in stages I and II of the algorithm. The RMSE for a given time series is computed as
\begin{align}
    \text{RMSE}
    &=
    \sqrt{\frac{1}{N_\mathrm{m}} \sum_{i=1}^{N_\mathrm{int}} \left[
        \sum_{k=k_i^o}^{k_i^f} (\hat{f}_k - \hat{f}^\mathrm{d}_i - \omega \hat{f}^\mathrm{v})^2
    \right]}.
\end{align}
Additionally, we compute for each time-series in the TAS dataset a ``naive" RMSE, obtained by ignoring changepoints and assuming the dry friction coefficient to be constant along the whole time-series. The histograms of both metrics are shown in Figure~\ref{fig:rmse_compare}, together with a 2D histogram that relates both RMSEs. To take into account the effect of the noise on the error, we estimate its standard deviation using a short window of data and subtract it from the RMSE, resulting in what we call the excess RMSE.

While the excess RMSE obtained under the assumption of constant $f^d$ has an average value of $0.13$, it goes down to $0.0024$ when considering a variable dry friction coefficient. This considerable decrease reveals the importance of considering the jumps of the dry friction coefficient. The 2D histograms, on the other hand, show a lack of correlation between both errors. This suggests that no other significant phenomena were present in the data that were not accounted for, as it would have produced a similar increase in RMSE with single and variable dry friction coefficients. 

Although the RMSE obtained after accounting for changepoints is small enough, considering that the minimum friction observed in the data is around $1.0$, it is higher than what we would expect if the noise were the only source of error. If such were the case, we would expect a mean excess RMSE of $0$ with a very small variance.

\begin{figure}[ht]
    \centering
    \includegraphics[width=\linewidth]{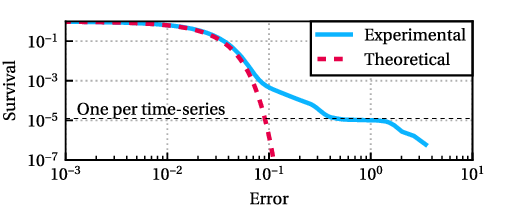}
    \caption{Survival function of the absolute error of the friction estimation for all measurements of all datapoints in the dataset, together with the theoretical distribution if random noise was the only cause of error.}
    \label{fig:error_survival}
\end{figure}

To investigate further, we collect the absolute error obtained for every measurement in the TAS dataset and generate a survival function, shown in Figure~\ref{fig:error_survival}. Assuming that Gaussian noise is the only source of error, we can compute the theoretical survival function using the $\chi^2$ distribution and compare it to the experimental one. Differences between the two distributions give information about errors due to incorrect estimations of the friction coefficients.

We observe a very close match between both functions until an error of approximately $0.08$. As the value of the experimental survival function is approximately $10^{-3}$, we can see that out of every $1000$ points, $999$ have an error that falls in the expected distribution, suggesting a good fit of the data. However, we observe many points with a higher error than expected, explaining the high mean RMSE observed in Figure~\ref{fig:rmse_compare}. The fact that the fraction of points with high error is so small suggests irregularities in the data and occasional estimation errors, rather than some fundamental issue in the model or the algorithm.

As in the previous subsection, there is a word of caution. Although a low RMSE is a good indicator of the correct fit of the coefficients, it is not a flawless metric. An algorithm that detects more changepoints than necessary will produce a reduced RMSE. In the following subsection we will explore the issue of false positives in more detail.

\subsection{Maximum Likelihood Assignment}

We analyze the assignment stage of the algorithm, tasked with determining which changepoints are caused by which FSS by running a dynamic programming algorithm to maximize the likelihood of the assignment.
As with the case of the changepoints, we do not have a ground truth of what the correct assignment of changepoints is. Instead, we will focus on two elements of the algorithm: the number of changepoints rejected, and the number of iterations required before an assignment is obtained.

Table~\ref{tab:rejections} shows the number of changepoints rejected as a histogram taken over all the datapoints in the TAS dataset. An unusually low number of rejections can be observed: 62\% of the datapoints are processed with no changepoints being rejected, while 96\% of the runs require three or fewer rejections. This shows that, while rejection is a needed component of the algorithm, it does not seem to be abused, considering that the average number of changepoints per run is around 26 (with a standard deviation of 10).


\begin{table}[ht]
    \centering
    \begin{tabular}{@{} c c c c c c c c c c c @{}}
        \toprule
         \textbf{$\#$ rejected} & $\mathbf{1}$ & $\mathbf{2}$ & $\mathbf{3}$ & $\mathbf{4}$ & $\mathbf{5}$ & $\mathbf{6}$ & $\boldsymbol{\geq}\mathbf{7}$
         \\
         \midrule
         TAS runs [$\%$] & $62.6$ & $18.3$ & $12.2$ & $4.5$ & $2.4$ & $0.5$ & $0.7$
         \\
         \bottomrule
         \addlinespace
    \end{tabular}
    \caption{Percentage of runs in the TAS dataset by number of changepoints rejected.}
    \label{tab:rejections}
\end{table}

\begin{figure}[ht]
    \centering
    \includegraphics[width=\linewidth]{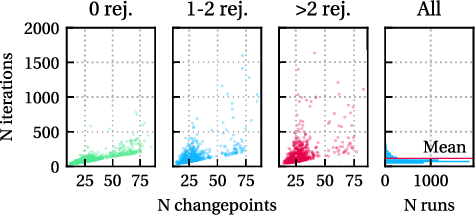}
    \caption{Relation between the number of iterations run by the assignment algorithm before obtaining the maximum likelihood assignment, and the number of changepoints to assign. A mean of $116.6$ iterations was obtained across all datapoints in the TAS dataset.}
    \label{fig:n_iterations}
\end{figure}

We are also interested in the computational cost of the algorithm, since it is well known that dynamic programming does not scale well with the number of states \cite{bertsekas_dynamic_2012}. So, it is essential to remember this when analyzing the algorithm. To do this, we focus on the number of iterations the algorithm takes to reach an optimal assignment, as it is proportional to the time it takes the algorithm to finish. In Figure~\ref{fig:n_iterations}, we show a scatter plot of runs from the TAS dataset by the number of changepoints they had, and the number of iterations it takes to compute their assignment. Additionally, we distinguish three groups according to the number of rejected changepoints: green for no rejections (accounting for $62\%$ of the points), blue for 1 to 3 rejections ($34\%$ of the points), and red for more than three rejections ($4\%$ of the points).

For the case of no rejections, we observe the number of iterations to scale roughly linearly with the number of changepoints. However, some outliers can be observed (likely coming from time-series in which several assignment hypotheses were kept until the end).
However, this linear scaling is not as straightforward when the number of rejected changepoints increases. In this situation, the relation between the number of changepoints and the number of iterations becomes less predictable: at best, it stays linear, while at worst, it gets exponential. This randomness is suspected to be due to where the rejected changepoint is located. A false changepoint may be easy to detect, or require ruling out many alternative hypotheses, notably increasing the number of iterations.

This points out the main limitation of rejection of changepoints, as it makes the assignment problem much more expensive. We should, therefore, be careful with the Average Run Length set for the changepoint detection, and not expect the assignment to take care of all false positives.

As a final note, we must keep in mind that, on average, the number of iterations is relatively small, around 115 iterations. This is because, although this number can reach high values when the number of changepoints and/or rejections is high, both numbers are relatively small for most runs.

\subsection{Anomaly Classification}

We present the paper's main results now, namely, the algorithm's accuracy in determining the anomaly affecting the RWA given a time series of measurements of spin rate and friction torque captured by the onboard sensors. Since the TAS dataset had a ground truth of the anomaly affecting the simulated RWAs used to generate the datasets, we can easily assess how well the algorithm performs the detection and classification of the anomaly.

Note that this step represents the combined work of every previous algorithm step. Hence, a good accuracy is itself an indicator of the capability of the complete algorithm to analyze the onboard data and determine the anomaly or anomalies affecting the RWA. In addition to the accuracy of the diagnosis, we study the impact of the number of bins used in the histogram generation on anomalies related to the FSS.

To determine the algorithm's accuracy at determining the anomaly present in an RWA, we compare its output to the labels in the TAS dataset. Although the algorithm can determine several anomalies at once, time-series in the dataset only have one anomaly at most. We consider the five possible labeled statuses (nominal + 4 anomalies) and determine, for each, how likely each anomaly is to be detected by the algorithm.

\begin{figure}[ht]
    \centering
    \includegraphics[width=\linewidth]{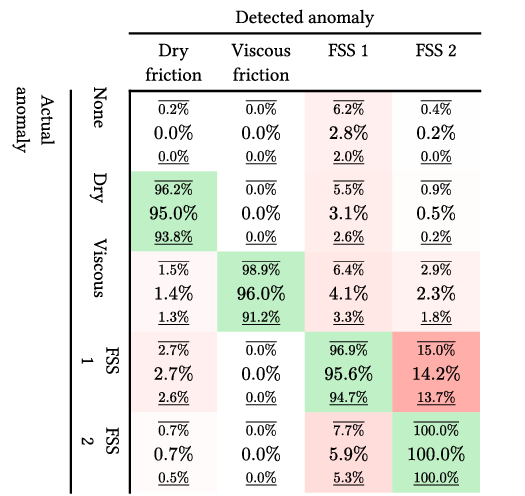}
    \caption{Anomaly detection probabilities obtained with 30 random splits of training and validation sets. The top number is the highest, the middle is the mean, and the bottom is the lowest among the 30 tests. Note that percentages do not add up to 100\%, since the same time-series can be classified with multiple anomalies simultaneously.}
    \label{fig:accuracy}
\end{figure}

The classification step involves training an SVM, done by splitting the dataset into a training set (20\% of the datapoints) and a validation set (80\% of the datapoints). Since this introduces some randomness in the process, we run the training and validation 30 times and show the minimum, mean, and maximum values to verify the stability of the classification. The results are shown in Figure~\ref{fig:accuracy}.

For the four anomalies considered, the probability of them being correctly detected is around $95\%$. The only exception is the FSS 2, which has an accuracy of $100\%$, which implies that a separating plane exists between the histograms of the friction values of the FSS 2 with and without the anomaly.

Another point of consideration is the probability that a datapoint is misclassified. These probabilities are generally low, with the highest number found in the probability that an anomaly in FSS 2 is detected for a datapoint with an anomaly in FSS 1, at $14\%$. This can be attributed to some confusion in the assignment process, which assigns a changepoint produced by one FSS to a different one.

Finally, we observe that the differences between the lowest and highest values obtained in different trainings are very small, with the most significant variation around $5\%$. This can be attributed to the simplicity of the SVM, as a small number of tunable hyperparameters reduces the risk of overfitting.

\subsection{Computation Time}

We now evaluate the times it takes the algorithm to run on the datapoints of the dataset provided by TAS, each containing around 80000 measurements of spin rate and friction. The tests were run on a laptop with an Intel® Core™ Ultra 7 165U. 

\begin{figure}[ht]
    \centering
    \includegraphics[width=\linewidth]{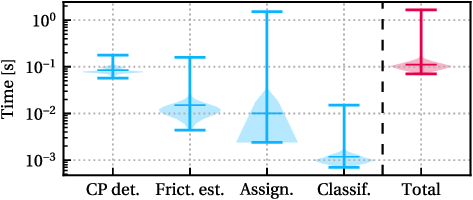}
    \caption{Distribution of computation time taken by each step of the algorithm, and by the complete algorithm. Each violin plot has 3 horizontal lines corresponding to the maximum, the mean, and the minimum times.}
    \label{fig:computimes}
\end{figure}

We first analyze the time taken by each step of the algorithm to run, together with the total computation time. This results in Figure~\ref{fig:computimes}, a violin plot with logarithmic scale to represent the large differences in computation time of the different steps.

The classification step takes less than 10 ms to run, which makes its contribution negligible. This is to be expected: generating the histograms is relatively cheap, and running an SVM involves little more than an inner product, which makes this step the cheapest to run, with a mean and maximum times of 1 and 10 ms respectively (the variations is due to a higher number of changepoints leading to more values to be included in the histograms).
The changepoint detection is the most expensive step to run, with an average runtime of 100 ms, but its cost has little variation as the computation of the LLR is done using vector operations.

The assignment component exhibits the most interesting behavior: it is fairly cheap on average (around 30 ms to run), but can reach up to 3 seconds in the dataset tested. 
The observations done on Figure~\ref{fig:n_iterations} suggest that such increases are related to a high number of changepoints rejected, which were observed to cause an exponential increase in the number of iterations of the algorithm. Although such an increase is very negative for the performance of the whole algorithm, only a very small percentage (around 5\%) of the datapoints had more than 3 rejected changepoints. In any cases, the computational times are very reasonable given the large number of measurements that each time-series contains, which makes the algorithm suitable to be deployed in the satellite.

\section{Conclusion}
We presented an algorithm to determine the anomaly present in an RWA from a window of measurements of spin rate and friction torque. The algorithm has 4 steps in order to address the peculiarities of the data, dealing with the need to detect changepoints and the properties of each anomaly shown exclusively through a labeled dataset. The results show an overall satisfactory performance of the algorithm, with an accuracy in the detection of anomalies of around $95\%$. Although the algorithm was developod with the TAS use case in mind, it can be applied to other friction systems that show a jumping behavior in their dry friction coefficient.

The analysis of the different steps of the algorithm provides an insight into possible future research on the topic. For instance, the changepoint algorithm could be improved using variable window sizes instead of fixed, although this would come at an additional computational cost. The assignment algorithm, although functional, can be improved using an improved algorithm developed for dynamic programming. Another aspect to consider is the need to know the transition functions of the FSS ($P_s^\mathrm{t}$) in order to run the algorithm: automatic methods for identifying them would make the algorithm easier to apply to new systems, without the need to manually identify them.




\bibliographystyle{IEEEtranCustom.bst} 
\bibliography{ultimate-journal}

\begin{IEEEbiographynophoto}{Alejandro Penacho Riveiros}
received the Engineering Degree in aerospace engineering from Universidad Carlos III de Madrid, in 2020, and the Civilingenjörsexamen degree in areospace engineering, with specialization in system engineering, from the KTH Royal Institute of Technology, Stockholm, Sweden, in 2023. He is currently pursuing a Ph.D. under the supervision of Karl H. Johansson and Matthieu Barreau in KTH Royal Institute of Technology, with focus in fault and anomaly detection and physics-informed machine learning.
\end{IEEEbiographynophoto}

\begin{IEEEbiographynophoto}{Nicola Bastianello} (Member, IEEE)
is a post-doc at the School of Electrical Engineering and Computer Science, and Digital Futures, KTH Royal Institute of Technology, Sweden. From 2021 to 2022 he was a post-doc at the Department of Information Engineering (DEI), University of Padova, Italy. He received a Ph.D. in Information Engineering at the University of Padova, Italy in 2021. During the Ph.D. he was a visiting student at the Department of Electrical, Computer, and Energy Engineering (ECEE), University of Colorado Boulder, Colorado, USA. He received a master's degree in Automation Engineering (2018) and a bachelor's degree in Information Engineering (2015) from the University of Padova, Italy. He currently serves in the IEEE CSS and EUCA Conference Editorial Boards. His research lies at the intersection of optimization and learning, with a focus on multi-agent systems.
\end{IEEEbiographynophoto}

\begin{IEEEbiographynophoto}{Karl H. Johansson} is Swedish Research Council Distinguished Professor in Electrical Engineering and Computer Science at KTH Royal Institute of Technology in Sweden and Founding Director of Digital Futures. He earned his MSc degree in Electrical Engineering and PhD in Automatic Control from Lund University. He has held visiting positions at UC Berkeley, Caltech, NTU and other prestigious institutions. His research interests focus on networked control systems and cyber-physical systems with applications in transportation, energy, and automation networks. For his scientific contributions, he has received numerous best paper awards and various distinctions from IEEE, IFAC, and other organizations. He has been awarded Distinguished Professor by the Swedish Research Council, Wallenberg Scholar by the Knut and Alice Wallenberg Foundation, Future Research Leader by the Swedish Foundation for Strategic Research. He has also received the triennial IFAC Young Author Prize and IEEE CSS Distinguished Lecturer. He is the recipient of the 2024 IEEE CSS Hendrik W. Bode Lecture Prize. His extensive service to the academic community includes being President of the European Control Association, IEEE CSS Vice President Diversity, Outreach \& Development, and Member of IEEE CSS Board of Governors and IFAC Council. He has served on the editorial boards of Automatica, IEEE TAC, IEEE TCNS and many other journals. He has also been a member of the Swedish Scientific Council for Natural Sciences and Engineering Sciences. He is Fellow of both the IEEE and the Royal Swedish Academy of Engineering Sciences.  
\end{IEEEbiographynophoto}

\begin{IEEEbiographynophoto}{Matthieu Barreau}
is an Assistant Professor in the Division of Decision and Control Systems at KTH Royal Institute of Technology, Stockholm, where he has been a faculty member since September 2023. He received his Ph.D. in Control Systems from LAAS-CNRS, Toulouse, in 2019, where his research focused on the stability analysis of coupled ordinary differential systems with string equations. His current research interests include physics-informed machine learning, traffic flow theory, infinite-dimensional systems, and controller synthesis.
He also holds a Master’s degree in Space Engineering from KTH Royal Institute of Technology, completed in 2016, and an Engineering degree in Aeronautical Engineering from ISAE-ENSICA, Toulouse, obtained in 2016. Prior to his current position, he served as an R\&D Manager at Tobii AB and as a Postdoctoral Researcher at KTH.
He has been recognized for his contributions to the field, notably receiving the Best French Ph.D. thesis award from GdR MACS and Club EEA in 2020.
\end{IEEEbiographynophoto}

\vfill

\end{document}